# Competing Antiferromagnetic-Ferromagnetic States in $d^7$ Kitaev Honeycomb Magnet

Hector K. Vivanco[1,2], Benjamin A. Trump[3], Craig M. Brown[3,4], and Tyrel M. McQueen[*,1,2,5]

[1]Department of Chemistry, The Johns Hopkins University, Baltimore, MD, 21218

[2]Institute for Quantum Matter, Department of Physics and Astronomy, The Johns Hopkins University, Baltimore, MD 21218

[3]NIST Center for Neutron Research, National Institute of Standards and Technology, Gaithersburg, MD 20899

[4]Department of Chemical and Biomolecular Engineering, University of Delaware, Newark, DE 19716

[5]Department of Materials Science and Engineering, The Johns Hopkins University, Baltimore, MD 21218

**Abstract**

The Kitaev model is a rare example of an analytically solvable and physically instantiable Hamiltonian yielding a topological quantum spin liquid ground state. Here we report signatures of Kitaev spin liquid physics in the honeycomb magnet $Li_3Co_2SbO_6$, built of high-spin $d^7$ ($Co^{2+}$) ions, in contrast to the more typical low-spin $d^5$ electron configurations in the presence of large spin-orbit coupling. Neutron powder diffraction measurements, heat capacity, and magnetization studies support the development of a long-range antiferromagnetic order space group of $C_C2/m$, below $T_N$ = 11 K at $\mu_0 H$ = 0 T. The magnetic entropy recovered between $T$ = 2 K and 50 K is estimated to be 0.6Rln2, in good agreement with the value expected for systems close to a Kitaev quantum spin liquid state. The temperature-dependent magnetic order parameter demonstrates a β value of 0.19(3), consistent with XY anisotropy and in-plane ordering, with Ising-like interactions between layers. Further, we observe a spin-flop driven crossover to ferromagnetic order with space group of $C2/m$ under an applied magnetic field of $\mu_0 H \approx 0.7$ T at $T$ = 2 K. Magnetic structure analysis demonstrates these magnetic states are competing at finite applied magnetic fields even below the spin-flop transition. Both the $d^7$ compass model, a quantitative comparison of the specific heat of $Li_3Co_2SbO_6$, and related honeycomb cobaltates to the anisotropic Kitaev model further support proximity to a Kitaev spin liquid state. This material demonstrates the rich playground of high-spin $d^7$ systems for spin liquid candidates, and complements known $d^5$ Ir- and Ru-based materials.

[*] mcqueen@jhu.edu

## Introduction

The quantum spin liquid (QSL) is an enigmatic and difficult to obtain state of matter, originally envisioned as a potential magnetic ground state for the S = 1/2 Heisenberg antiferromagnet on a triangular lattice as proposed by Anderson [1]. Strong QSL candidates have been proposed in the organics κ-(BEDT-TTF)$_2$Cu$_2$(CN)$_3$ and EtMe$_3$Sb[Pd(dmit)$_2$]$_2$, in addition to many minerals and related compounds based on the kagomé lattice, such as ZnCu$_3$(OH)$_6$Cl$_2$, ZnCu$_3$(OH)$_6$BrF, and BaCu$_3$V$_2$O$_8$(OH)$_2$ [2-5]. More recently, Alexei Kitaev provided an exactly solvable Hamiltonian with a QSL ground state based on bond-dependent anisotropic Ising interactions, realizable in real materials with a honeycomb lattice [6]. For a compound to follow the pure Kitaev model, there has to be an exact cancellation of the Heisenberg interactions between the magnetic ions in a honeycomb structure. Some Kitaev QSL candidates that have been examined extensively include Na$_2$IrO$_3$, Li$_2$IrO$_3$, H$_3$LiIr$_2$O$_6$, and α-RuCl$_3$ [7-10]. However, due to the bonding geometries of these materials, there are additional Heisenberg interactions giving rise to the Kitaev-Heisenberg model.

The focus for Kitaev QSLs has been primarily on low-spin $d^5$ materials. Recently, two concurrent theory papers provided support for a rich magnetic phase diagram for high-spin $d^7$ materials which includes the sought-after Kitaev QSL state [11,12]. The desired properties of a candidate are a pseudospin-1/2 compound with the transition metal (M) and ligand (L) having d-p-d orbital interactions in an octahedral environment. Based on the quantum compass model, an ideal bonding geometry for M-L-M of 90° leads to dominating Kitaev interactions [13]. This would potentially result in the spin liquid state. Certain delafossite compounds, such as Na$_2$Co$_2$TeO$_6$ and Na$_3$Co$_2$SbO$_6$, both of which have nearly ideal M-L-M bond angles, were predicted to be in proximity to the Kitaev QSL state.

Measurements of polycrystalline samples of Na$_2$Co$_2$TeO$_6$ and Na$_3$Co$_2$SbO$_6$ revealed zig-zag magnetic structures at low temperatures as well as signs of metamagnetic transitions under applied fields [14-17]. Another related compound, Ag$_3$Co$_2$SbO$_6$, displays a magnetic phase diagram composed of two antiferromagnetic (AFM) phases [18]. The single crystal investigations of the two proposed candidates confirm strong anisotropy and distinguishable magnetic properties. Na$_3$Co$_2$SbO$_6$ possesses the same experimental zig-zag AFM structure at low temperatures but is expected to be competing with the Néel state based on calculations performed [19]. Na$_2$Co$_2$TeO$_6$ demonstrates ferrimagnetism, which can be attributed to a mixture of Néel and zig-zag order [20]. These compounds exhibit Co-O-Co bond angles above the ideal 90°, averaging between 92° and 94°, with signs of possible frustrated magnetism. The next step is to identify and actualize materials near the ideal geometry.

Li$_3$Co$_2$SbO$_6$ has previously been reported as adopting either a honeycomb or orthorhombic structure depending on synthetic conditions [21,22], with the reported bond geometries of the honeycomb polymorph closer to the ideal 90° angle, at 90° to 92°. Here we report a robust synthesis of phase pure, honeycomb Li$_3$Co$_2$SbO$_6$, and investigate its magnetic properties through a combination of specific heat, magnetization, and neutron powder diffraction (NPD) measurements. Magnetization studies indicate a spin-flop transition at $\mu_0H \approx 0.7$ T at $T = 2$ K. Heat capacity measurements exhibit a magnetic entropy that saturates at 0.6Rln2, close to the Kitaev value of 0.5Rln2, with a temperature dependence that agrees with the predictions of the anisotropic Kitaev model. The peak of magnetic specific heat is pushed up in temperature with an applied magnetic field above the spin-flop transition, indicating the presence of significant ferromagnetic (FM) interactions. NPD under field reveals competing AFM-FM magnetic structures below the spin-flop transition and an XY-plane ordering model. Using these measurements, we construct a magnetic phase diagram, and place $Li_3Co_2SbO_6$ close to the Kitaev spin liquid state on the compass model. Our results demonstrate the rich playground of high-spin $d^7$ systems as spin liquid candidates, and complement known $d^5$ Ir- and Ru-based materials in exhibiting significant Kitaev interactions.

**Experimental Section**

Polycrystalline powder of honeycomb $Li_3Co_2SbO_6$ was synthesized from $Li_2CO_3$ (Alfa Aesar[†], 99.998%), $Co_3O_4$ (NOAH, 99.5%), and $Sb_2O_3$ (NOAH, 99.9%). $Li_2CO_3$ was dried at 200°C overnight prior to use. The starting reagents were mixed and ground in the appropriate stoichiometric ratio with an extra 4 mol% Li to compensate for volatilization. The resulting mixture was pelletized and heated to 700 °C, 800 °C, and 1100 °C sequentially for 1 h each in air, similar to the synthesis of $Li_3Ni_2SbO_6$ [23]. The sample was reground and pelletized between heat treatments. The final heating at 1100 °C required the sample to be air quenched in a desiccator. The resulting material was light red-brown and sensitive to the atmosphere. $Li_3Zn_2SbO_6$ was synthesized with a high temperature solid state reaction previously described [24]. The purity of the samples was confirmed using Rietveld refinements of powder x-ray diffraction (PXRD) patterns, collected on a Bruker D8 Focus diffractometer with a LynxEye detector using Cu K$\alpha$ radiation.

Physical properties were measured using a Quantum Design Physical Property Measurement System (PPMS). Magnetic susceptibility and heat capacity measurements were collected from $T = 2$ K to 300 K. Magnetization measurements were performed at intermediate fields between $\mu_0H = 0$ T to 2 T with a temperature range of $T = 2$ K to 13 K.

NPD measurements were conducted at the high resolution neutron powder diffractometer, BT1, at the NIST Center for Neutron Research (NCNR), using 60' in-pile collimation and a Ge(311) monochromator, with a wavelength of 2.077 Å. Approximately 3 g of $Li_3Co_2SbO_6$ was loaded into a vanadium can with an inner diameter of 6 mm and sealed with an indium O-ring in a He environment, and mounted onto a closed-cycle cooling refrigerator for measurements from $T = 3.6$ K to 30 K. Additional measurements were conducted at magnetic fields ranging from $\mu_0H = 0$ T to 7 T at $T = 3.6$ K and 30 K. A room temperature scan using 60' in-pile collimation and a Ge(733) monochromator with a wavelength of 1.1968 Å was also measured. Rietveld refinements and magnetic structure analysis were performed using GSAS-II and the k-Subgroupsmag program [25,26].

---

[†] Certain commercial equipment, instruments, or materials are identified in this paper to foster understanding. Such identification does not imply recommendation or endorsement by the National Institute of Standards and Technology, nor does it imply that the materials or equipment are necessarily the best available for the purpose.

NPD was collected in three groups of measurements: $\mu_0 H = 0$ T measurements, applied field measurements, and a room temperature (RT) measurement. The $\mu_0 H = 0$ T measurements includes long scans at $T = 3$ K and 30 K for magnetic structure determination, along with shorter scans to extract the temperature-dependent magnetic order parameter. Field measurements were performed similarly for magnetic structure determination under several applied fields. RT measurement was used to confirm crystal structure and atomic positions.

**Results and Discussion**

Synthesis & Structure Refinement

Using our high temperature solid state method, the honeycomb-$Li_3Co_2SbO_6$ is formed and is phase pure within the limit of laboratory PXRD. Consistent with prior reports, the heating profile is very important to avoid forming the undesired orthorhombic phase [22]. Quenching improves the crystallinity of the sample. Due to differing synthetic routes, a structure refinement was completed using a high-Q room temperature NPD measurement to confirm structure and atomic positions presented in Table I. There was an initial assumption for no site mixing between any atoms. However, the refinement was improved with the addition of Li/Co site mixing as seen in the ion exchange synthesis [21]. Such mixing increases the structural entropy, consistent with it being thermodynamically stable only at elevated temperatures. The resulting space group is $C2/m$ with the stoichiometry of $Li_3(Li_{0.2}Co_{1.8})SbO_{5.9}$. There is an additional impurity phase that is not present in the laboratory PXRD pattern that is indexed as $LiOH·H_2O$ with a 1.3 wt% (Fig. 1). This is similar to the impurities forming on $LiCoO_2$ in air, such as $Li_2CO_3$ scale forming on the surface [27]. We will be referring to the honeycomb, delafossite structure as $Li_3Co_2SbO_6$ for the rest of the work.

The Li-delafossite demonstrates bonding geometry closer to the $d^7$ Kitaev theory compared to the Na-delafossite analog. $Li_3Co_2SbO_6$ contains Co-O-Co bond angles between 90° to 91°, where $Na_3Co_2SbO_6$ has bond angles around 93°. The Li-delafossite should result in magnetic behavior more proximate to the Kitaev regime compared to the Na-delafossite.

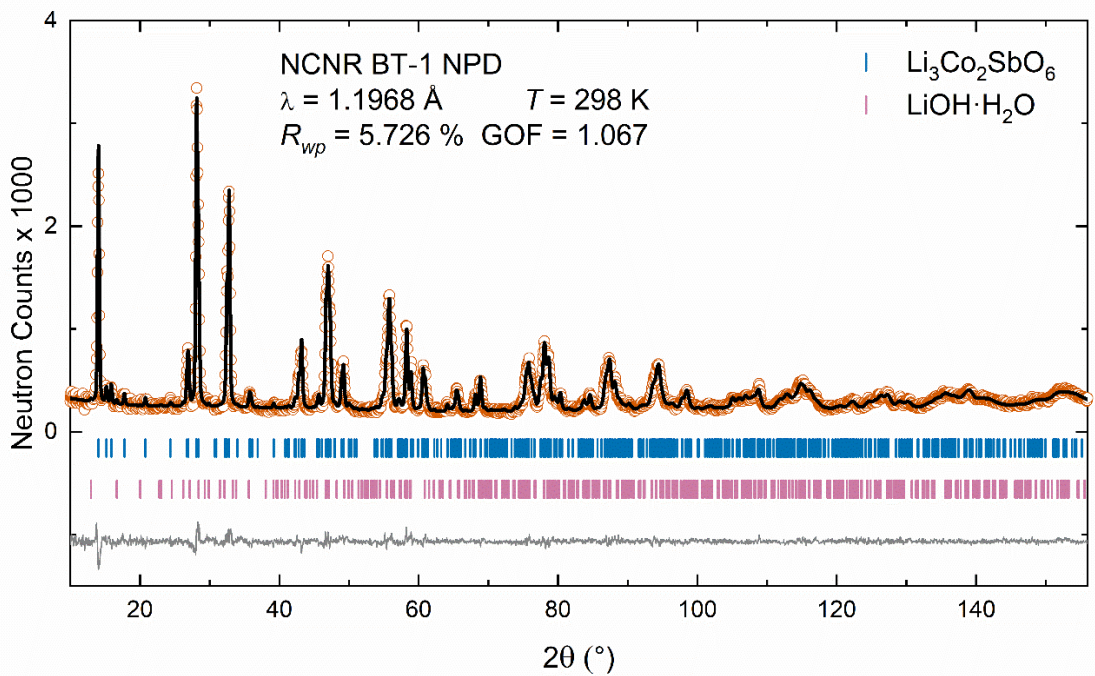

**Figure 1.** Rietveld refinement (black line) to high-resolution NPD pattern of Li$_3$Co$_2$SbO$_6$ collected at $T$ = 298 K (orange points), with difference curve (gray line). The blue ticks correspond to the Bragg reflections for the main phase and the purple ticks represent the LiOH·H$_2$O impurity with 1.3 wt%.

**Table I.** Unit cell parameters and fractional atomic coordinates for Li$_3$Co$_2$SbO$_6$ at $T$ = 298 K, from Rietveld refinement to a NPD Measurement with a wavelength of 1.1968 Å. Values in parentheses indicate one standard deviation obtained from GSAS-II refinement. The fit quality is given by a $R_{wp}$ = 5.726 % and a GOF = 1.067.

**Monoclinic *C*2/*m* Li$_{3.2}$Co$_{1.8}$SbO$_6$**

| $T$ (K) | $a$ (Å) | $b$ (Å) | $c$ (Å) | beta (°) | Volume (Å$^3$) |
|---|---|---|---|---|---|
| 298 K | 5.2300(14) | 9.0067(4) | 5.1834(13) | 110.179(4) | 229.177(12) |
| Atom | x | y | z | Occ. | $U_{iso}$ (Å$^2$) |
| Li1 | 0 | 0.1663(16) | 0.5 | 1 | 0.0217(21) |
| Li2 | 0 | 0.5 | 0.5 | 1 | 0.027(4) |
| Li_Co1 | 0 | 0.3377(10) | 0 | 0.1 | 0.0011(12) |
| Co1 | 0 | 0.3377(10) | 0 | 0.9 | 0.0011(12) |
| Sb | 0 | 0 | 0 | 1 | 0.0080(9) |
| O1 | 0.7620(7) | 0 | 0.2264(8) | 1 | 0.0086(7) |
| O2 | 0.2360(4) | 0.15698(26) | 0.2333(4) | 0.975* | 0.00688(31) |

* Fixed O2 occupancy for charge-balancing.

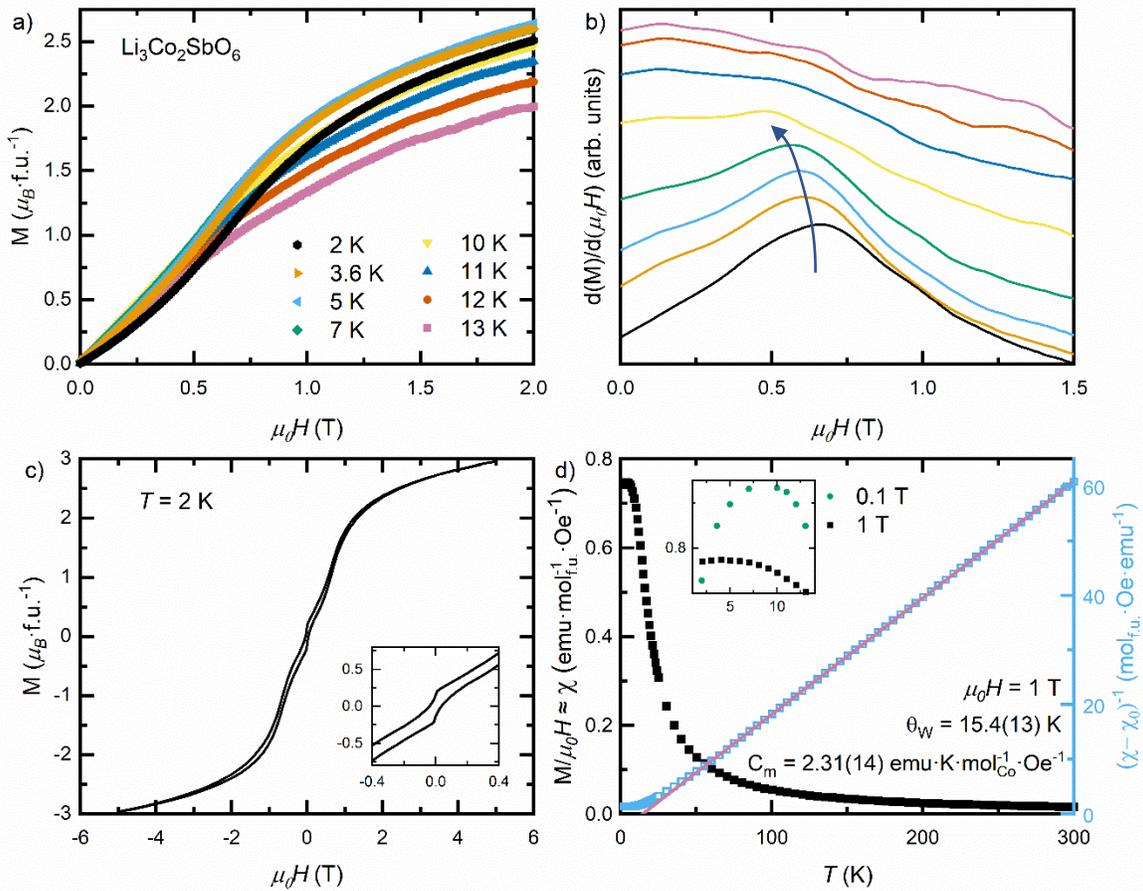

**Figure 2.** Magnetization isotherms (a) and the first derivative of magnetization (dM/d($\mu_0H$)) (b) for $Li_3Co_2SbO_6$ displaying curvature of metamagnetic behavior. A magnetization loop at $T$ = 2 K (c) demonstrates a small hysteresis indicative of FM interactions. Magnetic susceptibility, estimated as M/$\mu_0H$, at $\mu_0H$ = 1 T and the inverse susceptibility (d) with the Curie-Weiss fit (purple line) of the inverse susceptibility between $T$ = 60 K and 300 K.

Magnetic properties

Fig. 2a and 2b display magnetization curves and their first derivatives (dM/d($\mu_0H$)). Metamagnetic behavior for $Li_3Co_2SbO_6$ is apparent in the change in concavity of the M(H) curve, with a spin-flop transition at $\mu_0H \approx 0.7$ T for $T$ = 2 K. The spin-flop transition field decreases at elevated temperatures, consistent with both physical expectations and prior measurements [14-18,21,22]. A small hysteresis loop is observed in M(H) at $T$ = 2 K (Fig. 2c), analogous to the small hysteresis found in $Na_3Co_2SbO_6$. The presence of such hysteresis suggests dominating FM interactions [15]. Temperature-dependent magnetic susceptibility, estimated as $\chi$ = M/$\mu_0H$ and collected at $\mu_0H$ = 1 T (Fig. 2d), above the spin-flop transition field, indicates a rollover, consistent with the development of magnetic order, below $T \approx 10$ K, and follows Curie-Weiss behavior above

$T$ = 60 K. In contrast, the $\mu_0 H$ = 0.1 T magnetization measurement (Fig 2d. inset) demonstrates a sharp downturn after $T_N \approx$ 10 K. We find the high temperature susceptibility of the $\mu_0 H$ = 1 T measurement can be fit well assuming an effective magnetic moment of 4.30(3) $\mu_B$/Co$^{2+}$ and a Weiss temperature of $\theta_w$ = 15.4(13) K. These results are consistent with prior work, which have suggested values between 3.3 $\mu_B$/Co$^{2+}$ and 5.04 $\mu_B$/Co$^{2+}$ and $\theta_w$ = 14 K and 18.1 K, respectively [21,22].

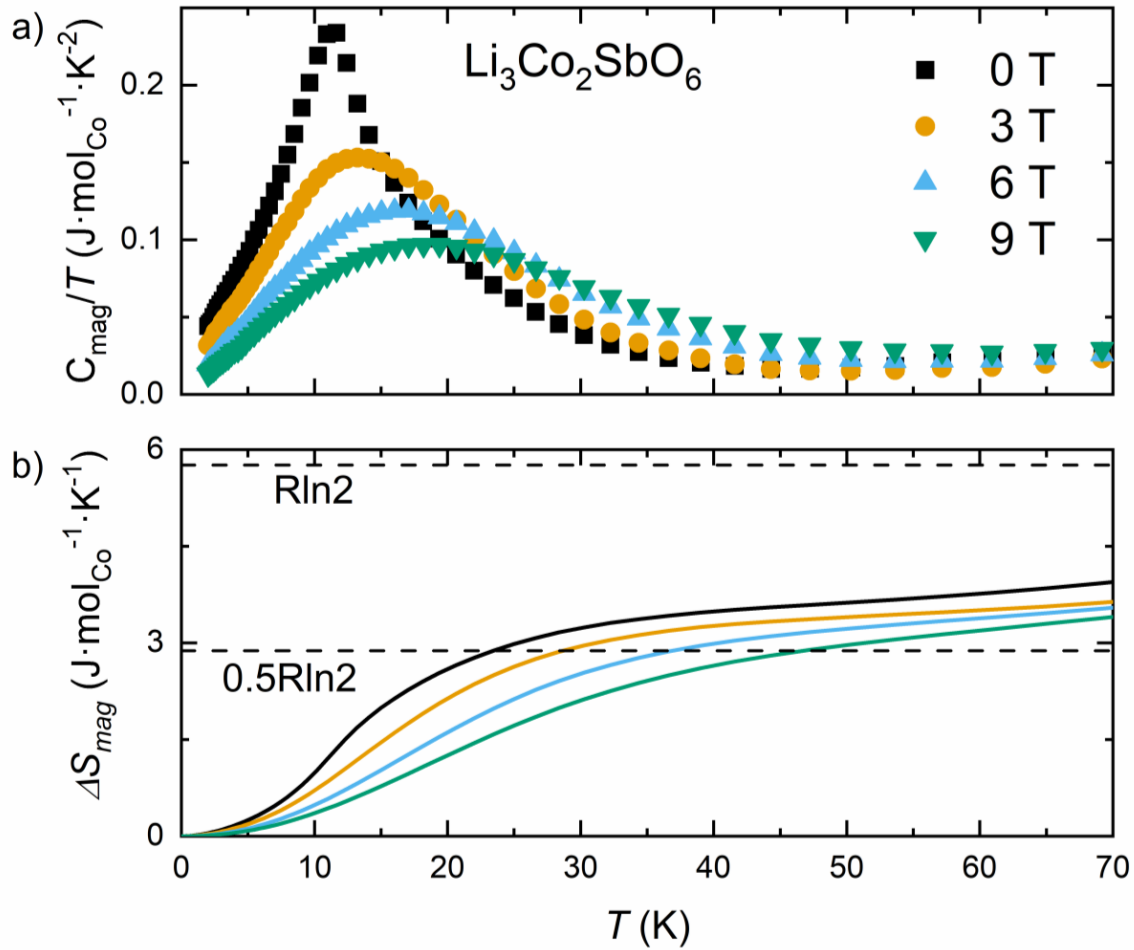

**Figure 3.** Temperature-dependent magnetic specific heat over temperature at several applied magnetic fields (a) demonstrates the transition increasing under applied fields behaving as a ferromagnet. Integrated magnetic entropy (b) plateaus at around 0.6Rln2 at $\mu_0 H$ = 0 T, approaching 0.5Rln2 under applied fields. This is in agreement with the anisotropic Kitaev model.

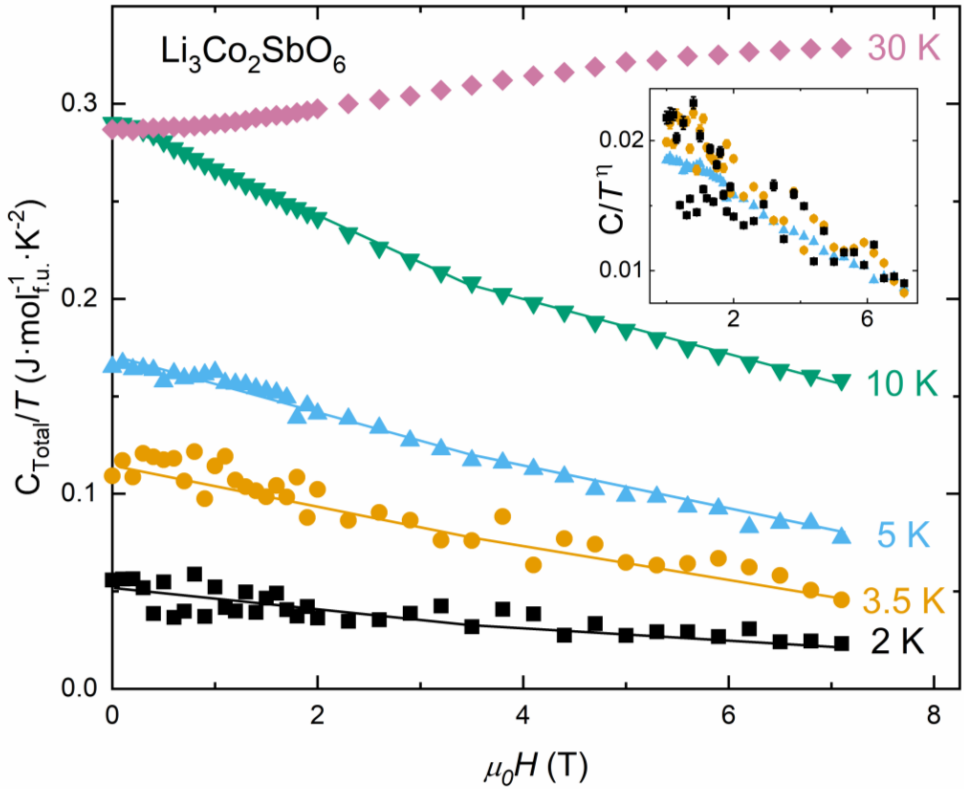

**Figure 4.** Field-dependent heat capacity at several temperatures (lines to guide the eye) with no λ-anomaly suggesting a spin-crossover transition. The inset presents overlapping isotherms that collapse onto a universal curve when heat capacity is divided by temperature to the power of 2.36 (η).

Heat Capacity Measurements

Heat capacity measurements were collected for $Li_3Co_2SbO_6$ as well as the nonmagnetic analog, $Li_3Zn_2SbO_6$, to subtract phonon contributions. There is an AFM transition present at $T \approx 11$ K with no applied field, consistent with low field magnetization measurements (Fig 2d. inset). The prominent λ-anomaly disappears under higher fields and develops into a broad transition that increases in temperature with field, behaving as a ferromagnet (Fig. 3a). Integration yields 3.6 J/mol$_{Co}$K ≈ 0.6Rln2 of magnetic entropy at $\mu_0 H = 0$ T. As the applied field increases, the recovered entropy up to $T = 50$ K decreases approaching 0.5Rln2 (Fig. 3b). This behavior is similar to a related cobaltate, $BaCo_2(P_{0.85}V_{0.15})_2O_8$, where the material saturates at 0.5Rln2 [28]. A saturation of recovered entropy of 0.5Rln2 matches the predictions of the Kitaev model presented by Singh et al. [29]. For isotropic S = 1/2 and anisotropic Kitaev models, there is an expected entropy plateau at $\Delta S_{mag} = 0.5\ln2$.

Field-dependent heat capacity measurements were also collected to determine the impact of the spin-flop on the bulk physical behavior, Fig. 4. There is no λ-anomaly, implying that the field driven transition is a spin-crossover, rather than a proper phase transition. Interestingly, we find the field dependent heat capacity at different temperatures can be collapsed onto a universal curve by scaling the heat capacity by temperature to the power of 2.36 (η). We have been unable to come up with a simple physical model to explain this scaling relationship, which should be a direction for future study [30].

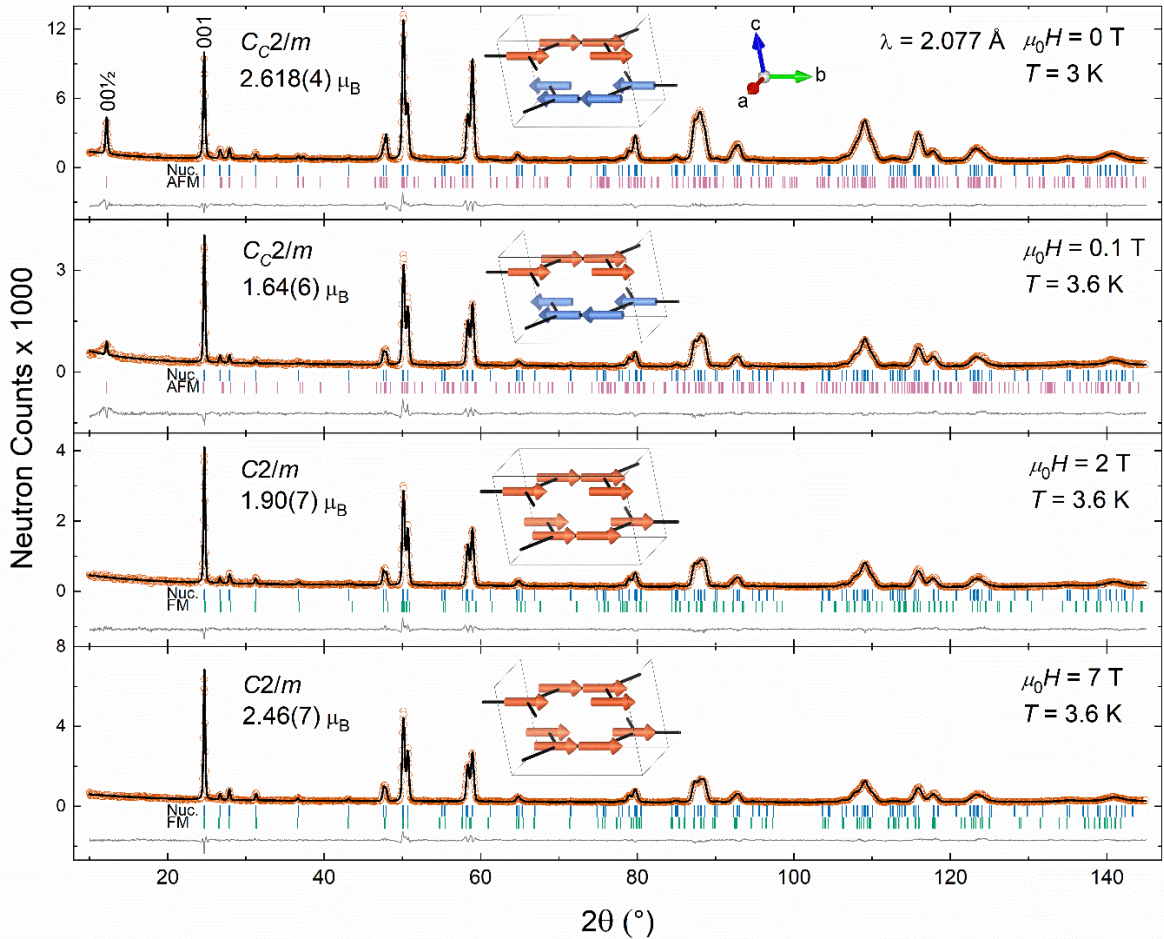

**Figure 5.** NPD of $Li_3Co_2SbO_6$ with Rietveld refinements to determine nuclear (Nuc.) and magnetic structures. At $\mu_0H$ = 0 T, the structure is AFM with alternating FM planes ($C_c2/m$). This AFM structure persists with reduced correlation length at $\mu_0H$ = 0.1 T. At higher fields, the magnetic structure is FM ($C2/m$).

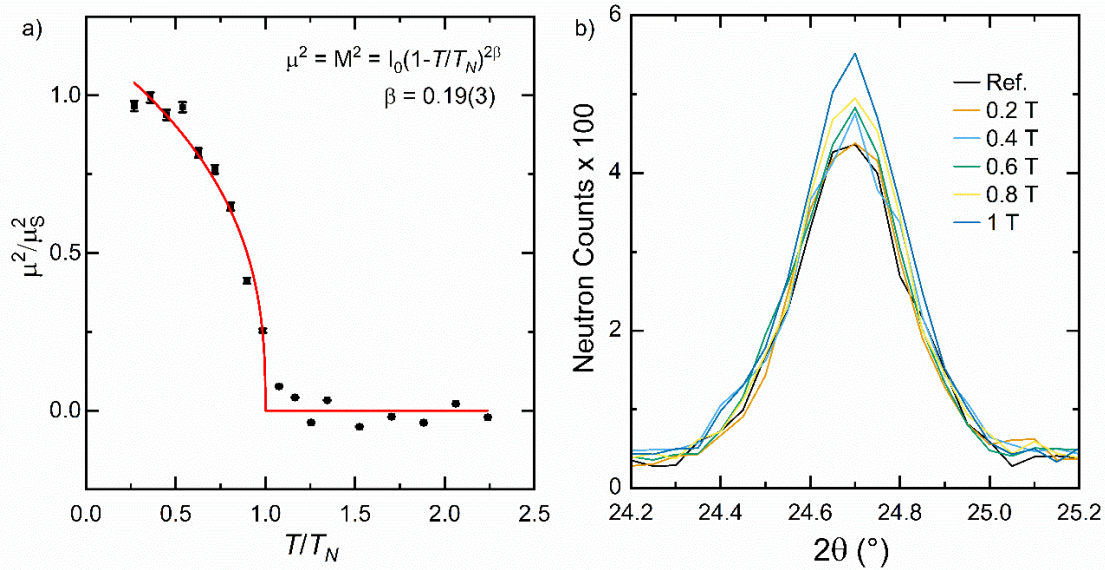

**Figure 6.** Temperature-dependent magnetic order parameter (a) at $\mu_0 H = 0$ T indicates an XY-plane ordering with Ising-like ordering between layers with a β value of 0.19(3). Power law analysis (red line) of integrated intensity and temperature for 2D XY-plane ordering is 0.23 while Ising ordering is 0.125. Field-dependent NPD scans of the (001) peak illustrate the FM peak developing above $\mu_0 H = 0.2$ T with a reference scan at $T = 30$ K (b). The peak increases in intensity compared to the reference below the spin-flop transition of $\mu_0 H = 0.7$ T which demonstrates the competing AFM-FM magnetic states.

Neutron Diffraction Measurements

To further understand temperature and field dependent states, more NPD was performed on $Li_3Co_2SbO_6$. NPD data at $T = 3$ K under no applied field displays an additional Bragg reflection corresponding to the AFM space group $C_c2/m$ (Fig. 5a). The refined magnetic moment of 2.63(4) $\mu_B$ is comparable to the moment of 2.53(3) $\mu_B$ at $T = 3$ K previously reported [22]. $Li_3Co_2SbO_6$ orders ferromagnetically in-plane with AFM coupling between layers, the first cobalt honeycomb delafossite to order in this manner. $CoTiO_3$ ($R\bar{3}$) orders in this same manner with a cobalt honeycomb [31]. The order parameter of the honeycomb $Li_3Co_2SbO_6$ (Fig. 6a) fits closest to an XY-plane ordering with an exponent β value of 0.19(3), consistent with the magnetic structure in-plane. This is in agreement with the order parameter determined for $Cu_3Co_2SbO_6$, where the order parameter fit to XY-plane ordering as well with the expected value of 0.23 [32]. The reduced exponent value for $Li_3Co_2SbO_6$ could be due to anisotropy of a powder measurement or due to more Ising-like ordering between layers [33], especially considering that the Li-delafossite has a smaller distance between layers compared to the Cu-delafossite, allowing for stronger interactions between layers.

Field-dependent measurements show rapid changes even under small applied magnetic fields. The $\mu_0 H = 0.1$ T measurement at $T = 3.6$ K demonstrates the reduction and broadening of the (00½) AFM peak resulting in an effective moment of 1.64(6) $\mu_B$ (Fig. 5b). The broadening of the AFM peak is consistent with the development of magnetic stacking faults and a reduction in AFM correlation length with an applied magnetic field. The field measurements at $\mu_0 H = 2$ T and 7 T support the FM magnetic structure of $C2/m$ [22]. The magnetic moment increases with field from 1.93(6) $\mu_B$ to 2.52(5) $\mu_B$ for $\mu_0 H = 2$ T and 7 T, respectively (Fig. 5c and 5d). To our knowledge, this is the first time this FM structure has ever been observed in a cobalt delafossite.

To elucidate more precisely the magnetic field driven changes in the magnetic order across the spin-flop transition observed at $\mu_0 H \approx 0.7$ T by magnetization measurements, the $T = 3.6$ K field-dependent scans were appropriately scaled and the non-magnetic contributions removed by subtraction of $T = 30$ K reference scans (Fig. 6b). These data demonstrate an increase in the (001) FM peak as the magnetic field increases to $\mu_0 H = 0.4$ T as the (00½) AFM peak continues to broaden and decrease in intensity. This corresponds to a FM structure competing with the AFM state of this Co-delafossite below the spin-flop transition. The (00½) AFM peak completely disappears at $\mu_0 H = 0.8$ T. This low temperature magnetic state analysis elucidates how $Li_3Co_2SbO_6$ is near the Kitaev QSL regime, as predicted by Khaliullin et al. and Motome et al. [11,12].

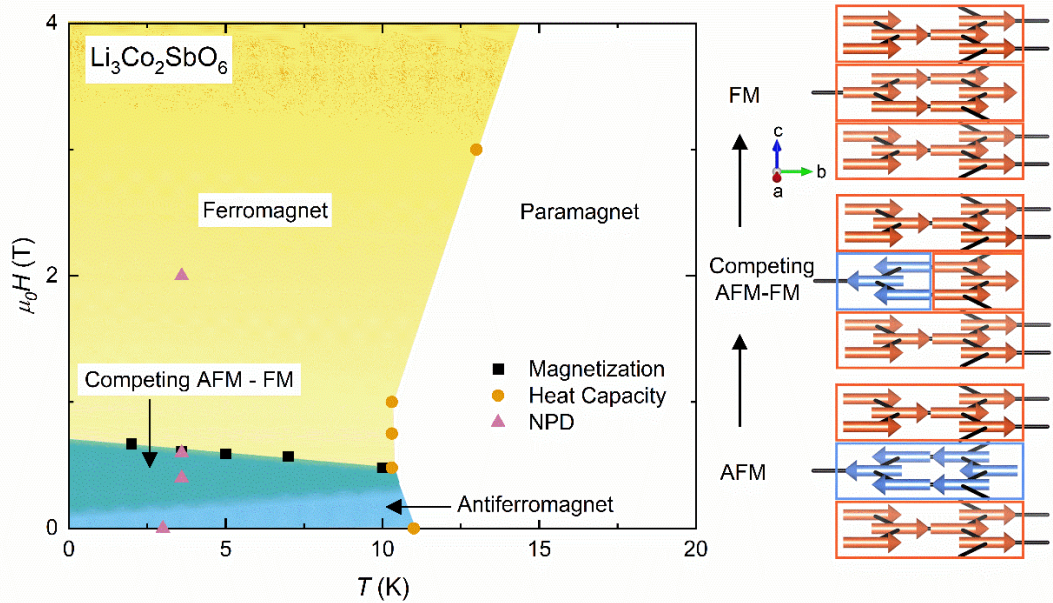

**Figure 7.** Magnetic phase diagram of $Li_3Co_2SbO_6$ presents the spin-crossover of AFM to FM states, with graphical illustrations of the magnetic states, under applied fields implying Kitaev interactions.

Phase Diagram

The magnetic phase diagram of $Li_3Co_2SbO_6$ is shown in Fig. 7. Heat capacity collected at $\mu_0H = 0$ T indicates magnetic ordering and NPD confirms the AFM state at $\mu_0H = 0$ T with FM planes that couple antiferromagnetically. Magnetization gives the phase boundary between the AFM and FM state with the spin-flop transition. In addition, the applied field NPD scans demonstrate the competing magnetic structures below the spin-flop transition. The competing behavior emerges around $\mu_0H = 0.4$ T. Higher field NPD measurements confirm a FM structure, in agreement with heat capacity behavior. Our magnetic phase diagram is unique in comparison to the other compounds within this delafossite family due to its competing magnetic structures [14-18]. The different behavior of $Li_3Co_2SbO_6$ can potentially be attributed to the bond angles being closer to the ideal 90° that are suspected to produce dominating Kitaev interactions. This places $Li_3Co_2SbO_6$ in the FM regime on the compass model adjacent to the Kitaev QSL state (Fig. 8b).

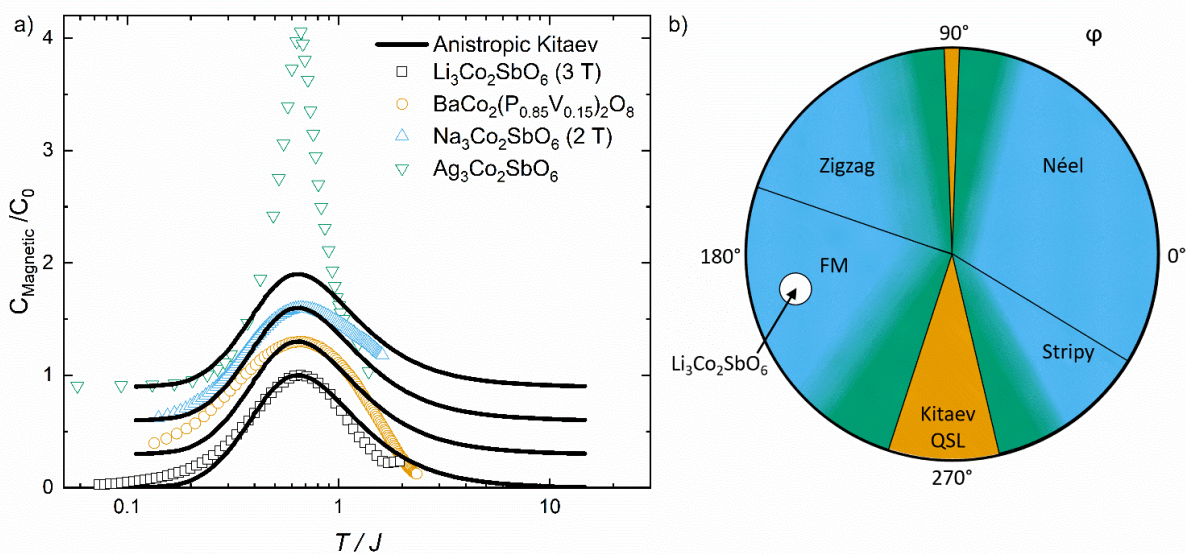

**Figure 8.** Normalized specific heat and temperature scaled by exchange interactions ($J$) with the calculated anisotropic S = 3/2 Kitaev model overlaid to display the comparison (a). Several high-spin $d^7$ compounds heat capacity data has been extracted to display Kitaev model signatures [18,19,28,29]. Each dataset is offset by 0.3 for clarity. The Kitaev compass model (b) with $Li_3Co_2SbO_6$ placed in the FM regime demonstrates how the Co-delafossites are proximate the Kitaev QSL state.

**Table II.** Extracted Exchange Interactions ($J$) based on anisotropic Kitaev Heat Capacity model.

| Compound Name | $J$ (K) | $C_0$ (J/mol$_{Co}$K) |
|---|---|---|
| $Li_3Co_2SbO_6$ (3T) | 27 | 2.41 |
| $BaCo_2(P_{0.85}V_{0.15})_2O_8$ | 13 | 1.82 |
| Na$_3$Co$_2$SbO$_6$ (2T) | 15 | 2.29 |
| Ag$_3$Co$_2$SbO$_6$ | 33 (28 [18]) | 2.70 |

Kitaev Compounds Comparison

Further indications of Kitaev physics in cobalt honeycombs comes from a more quantitative comparison of the specific heat to theoretical predictions. The anisotropic S = 3/2 Kitaev model with the highest anisotropic character is shown to have similar behavior in heat capacity (Fig. 8a) to both Li$_3$Co$_2$SbO$_6$ and related honeycomb cobaltates [29]. The recovered entropy of the S = 1/2 isotropic and all anisotropic models plateaus at 0.5ln2. The resemblance in broad heat capacity peak shape results in similar entropy recovered in these materials. The exchange interactions (*J*) have been extracted by scaling the heat capacity measurements to the model. Table II presents the *J* values with references to published *J* values. This provides an illustration of how the Co honeycombs are demonstrating Kitaev interactions.

Further support for this comes from the compass model (Fig. 8b): the FM regime (180° to 240°) is proximate the Kitaev QSL state (270°). The expected heat capacity behavior for the FM regime (180° to 210°) is a singular broad peak [34], in agreement with our observations.

**Conclusion**

Building on a new, robust synthesis for the honeycomb polymorph of Li$_3$Co$_2$SbO$_6$, we elucidate the presence of significant Kitaev interactions in this high-spin $d^7$ compound, and find a rich phase diagram, including a spin-flop transition at $\mu_0H \approx 0.7$ T and a spin-crossover to a FM state. Heat capacity demonstrates FM behavior under higher fields, which is confirmed using NPD to confirm a *C*2/*m* magnetic structure. Li$_3$Co$_2$SbO$_6$ is the first Co honeycomb delafossite investigated to display FM behavior, making it a singular comparator to α-RuCl$_3$ [35,36]. NPD illustrates competing antiferromagnet and ferromagnet magnetic behavior below the spin-flop transition but at finite magnetic field. The temperature-dependent magnetic order parameter indicates XY-plane ordering with the possibility of Ising-like ordering between layers. Based on the phase diagram of the high-spin $d^7$ model, Li$_3$Co$_2$SbO$_6$ is located proximate to the Kitaev QSL state. This Co honeycomb orders ferromagnetically under a small magnetic field and follows the Kitaev heat capacity and entropy model under an applied magnetic field. This family of high-spin $d^7$ compounds is a rich playground that would benefit from continued research, such as the application of pressure or strain to tune into the spin liquid regime [37,38]. More generally, our scaling analysis shows that multiple honeycomb cobaltates have specific heat behavior consistent with that expected under the anisotropic Kitaev model, implying a new era in the field of Kitaev physics, beyond low-spin $d^5$ compounds.

**Acknowledgments**

This work was supported as part of the Institute for Quantum Matter, an Energy Frontier Research Center funded by the U.S. Department of Energy, Office of Science, Office of Basic Energy Sciences, under Award DE-SC0019331. H.K.V. would like to thank Tom Halloran for useful discussions about neutron diffraction.

**References**

[1] P. W. Anderson, *Mater. Res. Bull.* **8**, 153-160 (1973).

[2] B. J. Powell and R. H. McKenzie, *Rep. Prog. Phys.* **74**, 5 (2011).

[3] M. R. Norman, *Rev. Mod. Phys.* **88**, 041002 (2016).

[4] C. M. Pasco, B. A. Trump, Thao T. Tran, Z. A. Kelly, C. Hoffmann, I. Heinmaa, R. Stern, and T. M. McQueen, *Phys. Rev. Mater.* **2**, 044406 (2018).

[5] R. H. Colman, F. Bert, D. Boldrin, A. D. Hillier, P. Manuel, P. Mendels, and A. S. Wills, *Phys. Rev. B* **83**, 180416(R) (2011).

[6] A. Kitaev, *Ann. Phys.* **321**, 2 (2006).

[7] Y. Singh and P. Gegenwart, *Phys. Rev. B* **82**, 064412 (2010).

[8] T. Takayama, A. Kato, R. Dinnebier, J. Nuss, H. Kono, L. S. I. Veiga, G. Fabbris, D. Haskel, and H. Takagi, *Phys. Rev. Lett.* **114**, 077202 (2015).

[9] K. Kitagawa, T. Takayama, Y. Matsumoto, A. Kato, R. Takano, Y. Kishimoto, S. Bette, R. Dinnebier, G. Jackeli, and H. Takagi, *Nature* **554**, 341-345 (2018).

[10] K. W. Plumb, J. P. Clancy, L. J. Sandilands, V. V. Shankar, Y. F. Hu, K. S. Burch, H.-Y. Kee, and Y.-J. Kim, *Phys. Rev. B* **90**, 041112(R) (2014).

[11] H. Liu and G. Khaliullin, *Phys. Rev. B* **97**, 014407 (2018).

[12] R. Sano, Y. Kato, and Y. Motome, *Phys. Rev. B* **97**, 014408 (2018).

[13] G. Jackeli and G. Khaliullin, *Phys. Rev. Lett.* **102**, 017205 (2009).

[14] L. Viciu, Q. Huang, E. Morosan, H. Zandbergen, N. Greenbaum, T. McQueen, and R. Cava, *J. Solid State Chem.* **180**, 1060 (2007).

[15] C. Wong, M. Avdeev, and C. D. Ling, *J. Solid State Chem.* **243**, 18 (2016).

[16] E. Lefrançois, M. Songvilay, J. Robert, G. Nataf, E. Jordan, L. Chaix, C. V. Colin, P. Lejay, A. Hadj-Azzem, R. Ballou, and V. Simonet, *Phys. Rev. B* **94**, 214416 (2016).

[17] A. K. Bera, S. M. Yusuf, A. Kumar, and C. Ritter, *Phys. Rev. B* **95**, 094424 (2017).

[18] E. A. Zvereva, M. I. Stratan, A. V. Ushakov, V. B. Nalbandyan, I. L. Shukaev, A. V. Silhanek, M. Abdel-Hafiez, S. V. Streltsov, and A. N. Vasiliev, *Dalton Trans.* **45**, 7373-7384 (2016).

[19] J.-Q. Yan, S. Okamoto, Y. Wu, Q. Zheng, H. D. Zhou, H. B. Cao, and M. A. McGuire, *Phys. Rev. Mater.* **3**, 074405 (2019).

[20] W. Yao and Y. Li, *Phys. Rev. B* **101**, 085120 (2020).

[21] M. I. Stratan, I. L. Shukaev, T. M. Vasilchikova, A. N. Vasiliev, A. N. Korshunov, A. I. Kurbakov, V. B. Nalbandyan, and E. A. Zvereva, *New J. Chem.* **43**, 13545 (2019).

[22] A. J. Brown, Q. Xia, M. Avdeev, B. J. Kennedy, and C. D. Ling, *Inorg. Chem.* **58**, 13881-13891 (2019).

[23] E. A. Zvereva, M. A. Evstigneeva, V. B. Nalbandyan, O. A. Savelieva, S. A. Ibragimov, O. A. Volkova, L. I. Medvedeva, A. N. Vasiliev, R. Klingeler, and B. Buechner, *Dalton Trans.* **41**, 572-580 (2012).

[24] C. Greaves and S. M. A. Katib, *Mater. Res. Bull.* **25**, 1175-1182 (1990).

[25] B. H. Toby and R. B. Von Dreele, *J. Appl. Cryst.* **46**, 544-569 (2013).

[26] J. M. Perez-Mato, S. V. Gallego, E. S. Tasci, L. Elcoro, G. de la Flor, and M. I. Aroyo, *Annu. Rev. Mater. Res.* **45**, 217-248 (2015).

[27] L. Wang, B. Chen, J. Ma, G. Cui, and L. Chen, *Chem. Soc. Rev.* **47**, 6505-6602 (2018).

[28] R. Zhong, M. Chung, T. Kong, L. T. Nguyen, S. Lei, and R. J. Cava, *Phys. Rev. B* **98**, 220407(R) (2018).

[29] J. Oitmaa, A. Koga, and R. R. P. Singh, *Phys. Rev. B* **98**, 214404 (2018).

[30] I. Kimchi, J.P. Sheckelton, T.M. McQueen, and P.A. Lee, *Nat. Comm.* **9**, 4367 (2018).

[31] R. E. Newnham, J. H. Fang, and R. P. Santoro, *Acta Cryst.* **17**, 240 (1964).

[32] J. H. Roudebush, N. H. Andersen, R. Ramlau, V. O. Garlea, R. Toft-Petersen, P. Norby, R. Schneider, J. N. Hay, and R. J. Cava, *Inorg. Chem.* **52**, 6083-6095 (2013).

[33] A. Taroni, S. T. Bramwell, and P. C. W. Holdsworth, *J. Phys.: Condens. Matter* **20**, 275233 (2008).

[34] Y. Yamaji, T. Suzuki, T. Yamada, S. Suga, N. Kawashima, and M. Imada, *Phys. Rev. B* **93**, 174425 (2016).

[35] A. Koitzsch, E. Müller, M. Knupfer, B. Büchner, D. Nowak, A. Isaeva, T. Doert, M. Grüninger, S. Nishimoto, and J. van der Brink, *arxiv*: 1709.02712 (2017).

[36] Y. Tian, W. Gao, E. A. Henriksen, J. R. Chelikowsky, and L. Yang, *Nano Lett.* **19**, 7673-7680 (2019).

[37] H. Liu, J. Chaloupka, and G. Khaliullin, *arXiv*:2002.05441 (2020).

[38] Z. Wang, J. Guo, F. F. Tafti, A. Hegg, S. Sen, V. A. Sidorov, L. Wang, S. Cai, W. Yi, Y. Zhou, H. Wang, S. Zhang, K. Yang, A. Li, X. Li, Y. Li, J. Liu, Y. Shi, W. Ku, Q. Wu, R. J. Cava, and L. Sun, *Phys. Rev. B* **97**, 245149 (2018).